\documentclass[twocolumn,footinbib,showpacs,preprintnumbers,amsmath,amssymb,prl,aps,10pt]{revtex4-1}

\usepackage{dcolumn}
\usepackage{bm}
\usepackage{units}
\usepackage{ifpdf}
\usepackage[dvips]{graphicx}  
	  
\begin{document}


\title{Trapping Ultracold Atoms in a Time-Averaged Adiabatic Potential}

\author{M. Gildemeister}
\author{E. Nugent}
\author{B. E. Sherlock}
\author{M. Kubasik}
\author{B. T. Sheard}
\author{C. J. Foot}
\affiliation{Clarendon Laboratory, University of Oxford, Parks Road, Oxford, OX1 3PU, United Kingdom}

\date{\today}

\begin{abstract}

We report the first experimental realization of ultracold atoms confined in a 
time-averaged, adiabatic potential (TAAP). This novel trapping technique involves 
using a slowly oscillating ($\sim$ kHz) bias field to time-average the instantaneous 
potential given by dressing a bare magnetic potential with a high frequency ($\sim$ MHz) 
magnetic field. The resultant potentials provide a convenient route to a variety of 
trapping geometries with tunable parameters.
We demonstrate the TAAP trap in a standard time-averaged orbiting 
potential trap with additional Helmholtz coils for the introduction of 
the radio frequency dressing field. We have evaporatively cooled 5 $\times 10^4$ atoms of $^{87}$Rb  
to quantum degeneracy and observed condensate lifetimes of over \unit[3]{s}. 

\end{abstract}

\pacs{37.10.Gh,32.80.Xx,03.75.-b}
\maketitle

The use of increasingly sophisticated magnetic potentials has been crucial to 
the evolution of cold atom research. The time-averaged adiabatic potential (TAAP) 
trap proposed in \cite{Lesanovsky2007} combines the techniques of time-averaging 
and radio frequency (RF) dressing to produce versatile potentials with a 
rich variety of different geometries. 
Since the RF dressing only couples electronic ground states of the atom the rate of relaxation by spontaneous 
processes is negligible. The potential is smooth with no small scale corrugations because the trapping region is 
located far from the field generating coils. As shown in \cite{Lesanovsky2007}, 
the TAAP can easily sculpt complex trapping geometries such as a double-well 
or ring trap, which can be 
adiabatically modified. Here we present an experimental realization 
of the particular case of a double-well TAAP and an efficient way of loading it from a 
time-orbiting potential (TOP) trap.

The technique of time-averaging involves the introduction of a time dependence to 
a static potential at a frequency higher than the atoms can respond to 
kinematically, but significantly lower than the local Larmor frequency. As 
a result atoms experience a modified 
potential whilst preserving their initial $m_{F}$ state.
The application of a suitable rotating bias field to a quadrupole potential can 
circumvent Majorana losses by ensuring the field zero orbits at a radius 
greater than the extent of the atom cloud.
For a bias field $B_{T}$ rotating at angular 
frequency $\omega_{T}$ the criterion for time-averaging can be stated as 
$\vert g_{F}\mu_{B}B_{T}/\hbar\vert\gg\omega_{T}>\omega_{r}$, where
$g_{F}$ is the Land\`{e} g-factor, $\mu_{B}$ the Bohr magneton, 
$\hbar$ the reduced Planck constant and $\omega_{r}$ the 
oscillation frequency of atoms in the time-averaged potential. 
This time-averaged, orbiting potential not only expedited 
the first realization of a Bose-Einstein Condensate (BEC) in a dilute alkali vapor 
\cite{Anderson1995}, but also paved the way for more sophisticated magnetic 
potentials such as a double-well \cite{Thomas2002} and ring-shaped potentials 
\cite{Arnold2006,Griffin2008,Gupta2005}. 
The general concept of using time-averaging to create novel shapes of confining potentials
has also been demonstrated by experiments on dipole
force traps \cite{Rudy2001,Friedman2000,Henderson2009}.

\begin{figure}[h]
\begin{center}
\includegraphics[scale = 1]{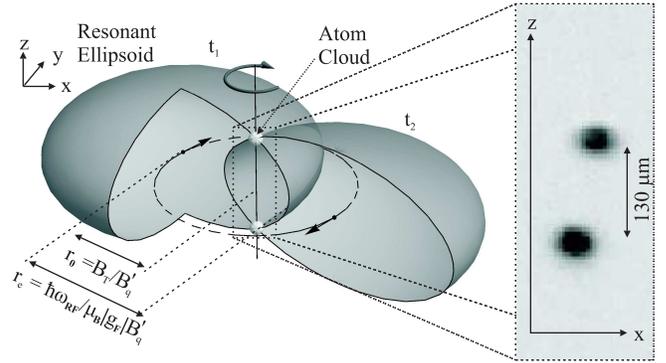}
\end{center}
\begin{flushleft}
\caption[Taap Trap Loading]{\label{fig:Loading} The instantaneous potential of the TAAP trap 
at two different times in the rotation cycle. The intersection of this potential with the 
rotation axis gives the minima of the TAAP at which the AP remains stationary. 
The inset shows an in-situ absorption image 
of a thermal cloud trapped at these positions. We attribute the tilt evident in 
this image to a misalignment of the rotation axis with respect to the symmetry 
axis of the quadrupole potential.}
\end{flushleft}
\end{figure} 

The repertoire of magnetic potentials was further extended by RF dressing, 
as proposed by Zobay and Garraway \cite{Zobay2001}.
Their technique brought about the possibility of 
highly asymmetric 2D magnetic potentials \cite{Colombe2004b, White2006, 
Morizot2007}. It has also been applied in atom chip experiments 
to give flexible double-well potentials  \cite{Schumm2005, Hofferberth2006} 
as described in detail elsewhere 
\cite{Zobay2001, Hofferberth2007, Lesanovsky2006a, Lesanovsky2006b}.
RF dressing involves a magnetic field oscillating at a frequency $\omega_{RF}$ comparable
to the atomic Larmor frequency which drives transitions 
between Zeeman sub-levels in a hyperfine manifold 
of the atom's electronic ground state. Here the eigenstates of the system 
are the dressed states \cite{Cohen1992}. The variation of the 
corresponding eigenenergies with position gives rise to an adiabatic 
potential (AP) given by 
\begin{equation}
\label{eqn:DressedStatePotentials} U_{AP}({\bf r}) = m_{F}\hbar
\sqrt{\delta^2({\bf r}) + \Omega_{R}^2({\bf r})},
\end{equation}
where $\delta({\bf r}) = \vert g_F  \mu_{B} B({\bf r})/\hbar\vert - \omega_{RF}$ is the angular frequency 
detuning and  
$\Omega_{R}({\bf r})$ the Rabi frequency given by 
\begin{equation}
\label{eqn:RabiCoupling}
\Omega_{R}({\bf r})= \vert\frac{g_F \mu_{B}}{2\hbar} \frac{{\bf B}({\bf r})}
{\vert {\bf B}({\bf r})\vert} \times {\bf B}_{RF} \vert. 
\end{equation}  
In this experiment we use a quadrupole field of the form
\begin{equation}
\label{eqn:Quadrupole}
{\bf B}_{q}({\bf r}) = B_{q}^{\prime} [x \hat{\bf e}_x + y
\hat{\bf e}_y - 2z \hat{\bf e}_z],
\end{equation}
where $B_{q}^{\prime}$ is the radial gradient.
When dressed with an RF field ${\bf B}_{RF}$ the upper dressed state has a 
minimum on an ellipsoidal iso-B surface where $\delta({\bm r}) =0$. 
The vectorial nature of the coupling 
(see Eq.\ \ref{eqn:RabiCoupling}) gives a variation 
of the potential on the ellipsoidal shell itself: for linearly 
polarized RF with ${\bf B}_{RF}(t) = B_{RF}\cos(\omega_{RF}t)
{\bf \hat{e}_{z}}$ the coupling varies from maximum on the equator to 
zero at the poles. For circularly polarized RF with ${\bf B}_{RF}(t) = B_{RF}[\cos(\omega_{RF}t)
{\bf \hat{e}_{x}} \pm \sin(\omega_{RF}t){\bf \hat{e}_{y}}]$ the coupling is maximum 
at one pole and zero on the other \footnote{In making the rotating wave approximation the time
dependence of the Rabi frequency is removed.}. 

\begin{figure}
\begin{center}
\includegraphics[scale = 1]{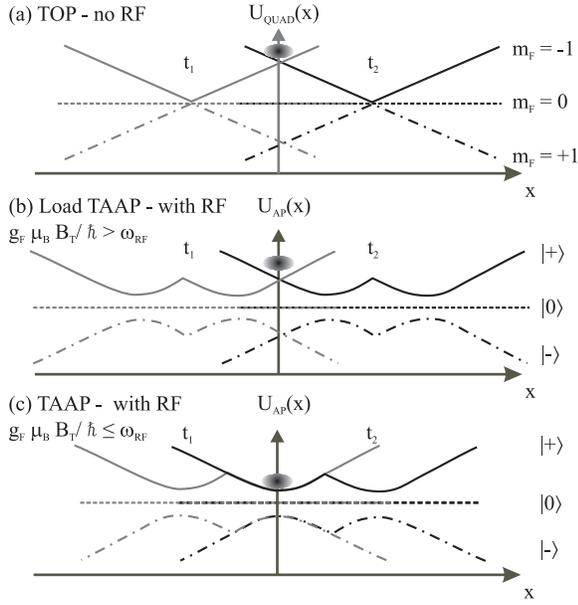}
\end{center}
\begin{flushleft}
\caption[Time Averaging]{\label{fig:TimeAveraging}The instantaneous potential of the 
(a) TOP, (b) TAAP loading and (c) TAAP potentials along the $x$ axis at time 
t$_{1}$ in the rotation cycle and time t$_{2}$ half a period later.}
\end{flushleft}
\end{figure}

RF dressed-state adiabatic potentials may themselves be time-averaged to 
give a new class of potentials referred to as time-averaged adiabatic potentials.
We have generated a double-well TAAP by applying RF radiation 
to a conventional TOP trap. The instantaneous potential of the 
TOP is a quadrupole field, in a TAAP trap this is dressed to give the 
ellipsoidal surface as described above. 
The oscillating bias field of the TOP trap, $ {\bf B}_{T}(t) = 
B_{T}[\cos(\omega_{T}t) {\bf \hat{e}_{x}} + \sin(\omega_{T}t) {\bf \hat{e}_{y}}]$, 
causes the ellipsoidal surface to orbit in the $xy$-plane 
about the axis of rotation as illustrated in Fig.\ \ref{fig:Loading}. 
When $\omega_{RF}>\vert g_{F}\mu_{B}B_{T}/\hbar \vert$ the ellipsoidal 
surface intersects the rotation axis at two points; these two 
points define the minima of the time-averaged potential.  

\begin{figure}[h]
\begin{center}
\includegraphics[scale = 1]{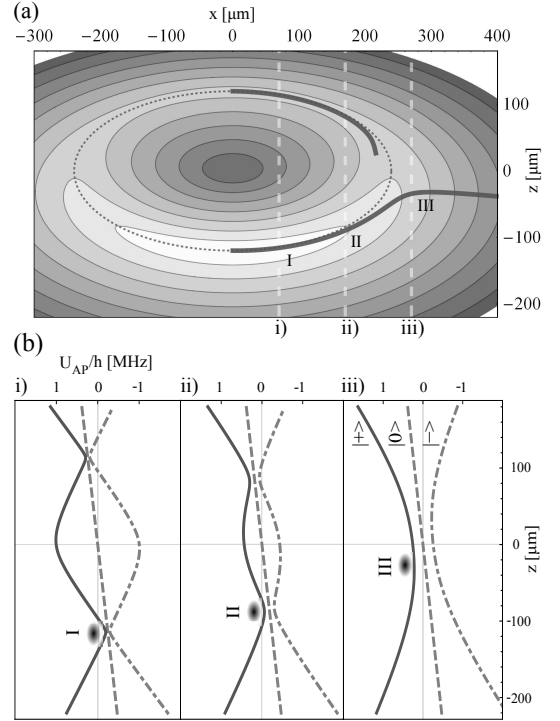}
\end{center}
\begin{flushleft}
\caption[Taap Trap Position Data]{\label{fig:ContourPlot} (a) A contour plot of the 
AP in the $xz$-plane for a gradient of $B_{q}^{\prime} =$ \unit[84]{G/cm} and 
for linearly polarized RF with $B_{RF} =$ \unit[0.5]{G}. The dashed lines indicate the position of the 
rotation axis of the ellipsoid for different values of $B_{T}$.\\ 
(b) Shows the potential along these lines for the three dressed states. The slope 
in the potentials corresponds to the gravitational potential energy of a 
$^{87}$Rb atom. Note the variation of the coupling at the atom's position.}
\end{flushleft}
\end{figure}

We load atoms into the double-well TAAP from a TOP using the following 
procedure. First, we prepare a sample of 
cold atoms in the TOP trap (see Fig.\ \ref{fig:TimeAveraging}(a)). 
The RF dressing field is then rapidly switched on while ensuring $B_{T}$
satisfies the inequality $\omega_{RF} < \vert g_{F}\mu_{B}B_{T}/\hbar\vert < 2\omega_{RF}$ (Fig.\ 
\ref{fig:TimeAveraging}(b)). The lower bound ensures the atoms are 
loaded into the correct dressed state while   
the upper bound is to prevent higher harmonics of $\omega_{RF}$ coming into 
resonance and causing unwanted evaporation. At this stage the modification 
of the time-averaged potential is minimal (the ellipsoids of Fig.\ \ref{fig:Loading} 
do not yet intersect with the rotation axis). Decreasing the TOP field to 
$\omega_{RF} = \vert g_{F}\mu_{B}B_{T}/\hbar \vert$ transfers the atoms onto the ellipsoidal 
shell (Fig.\ \ref{fig:TimeAveraging}(c)). A further decrease in the TOP field 
gives rise to a double-well potential in the $z$ direction. Note that the atoms 
stay on resonance at {\it all} times in the TAAP trap. The well separation 
is given by the distance between the points of intersection of the ellipsoidal surface 
with the rotation axis: a decrease in $B_{T}$ moves the atom clouds further along 
the ellipsoid thus increasing their separation. The TAAP potential in the 
$z$ direction is that along the rotation axis of the ellipsoid as depicted in Fig.\ \ref{fig:ContourPlot}. 
The effect of the time-averaging is to give confinement along the surface 
thus preventing the atoms spreading out over the ellipsoid. 

Our TOP trap (with $\omega_{T} = 2\pi \times$ \unit[7]{kHz}) apparatus routinely produces BEC's of 
1$\times 10^5$ atoms of $^{87}$Rb in the $\left|F=1,m_{F} =
-1\right>$ hyperfine state. 
The RF dressing field is applied through coaxial coil pairs placed
symmetrically about the trap center along the $x,y$ and $z$ axes;
this allows any arbitrary polarization but in these experiments it
was either polarized circularly with ${\bf B}_{RF}(t)$ in the $xy$-plane
or linearly with ${\bf B}_{RF}(t) \propto \bf \hat{e}_z$.   

\begin{figure}
\begin{center}
\includegraphics[scale = 1]{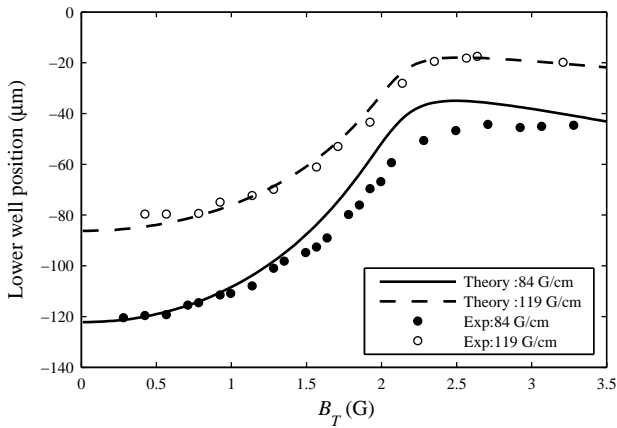}
\end{center}
\begin{flushleft}
\caption[Taap trap position data]{\label{fig:PositionData} Vertical displacement of the atoms from 
the static quadrupole field centre in the lower well of the 
TAAP trap as a function of the magnitude of the rotating bias field $B_{T}$. This is plotted for 
two different field gradients $B_{q}^{\prime} = $ \unit[84]{G/cm} and $B_{q}^{\prime} = $ \unit[119]{G/cm}. }
\end{flushleft}
\end{figure}

In Fig.\ \ref{fig:PositionData} the vertical position of atoms in the lower well of the TAAP trap 
is plotted as a function of the magnitude of the rotating bias field $B_{T}$ for two different 
quadrupole gradients. Here we apply linearly polarized RF along the $z$ direction with 
$B_{RF} =$ \unit[0.5]{G} and $\omega_{RF} = 2\pi \times$ \unit[1.4]{MHz}. For these parameters 
the ellipsoidal surface touches the rotation axis when $B_{T} = $ \unit[2]{G} at which 
stage the atoms are loaded into the TAAP 
trap. The polarization effects mentioned above and the gravitational sag of the atoms 
mean that as $B_{T}$ is lowered the atoms do not follow the perfect ellipsoidal trajectory  
that one would expect from the idealized picture above (in Fig.\ \ref{fig:ContourPlot} 
compare the dotted ellipsoid to the actual position of the minima shown in black).  

\begin{figure}
\begin{center}
\includegraphics[scale = 1]{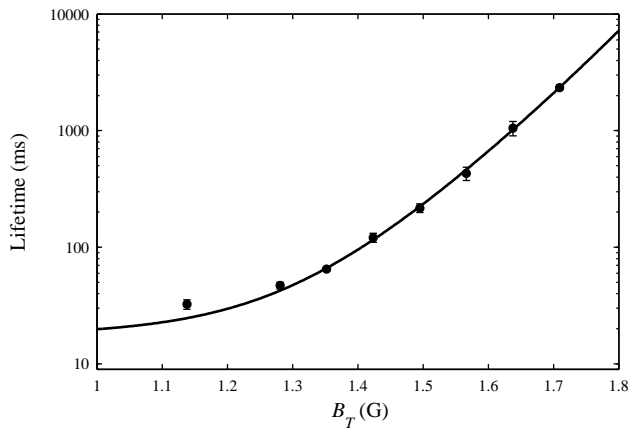}
\end{center}
\begin{flushleft}
\caption[Taap Trap Lifetime Data]{\label{fig:lifetime}The lifetime of atoms in the TAAP trap for linearly 
polarized RF with $B_{RF} =$ \unit[0.5]{G} and a radial quadrupole gradient of $B_{q}^{\prime} =$ 
\unit[84]{G/cm}. For this polarization of RF the $\Omega_{R}({\bf r})$ varies linearly with $B_{T}$ (see 
text for more details).}
\end{flushleft}
\end{figure}

It is evident in Fig.\ \ref{fig:ContourPlot} that gravitational sag makes it 
impossible to load into the upper well for this quadrupole gradient 
(as denoted by the discontinuity in the black line in Fig.\ \ref{fig:ContourPlot} indicating 
positions where there is no minimum in the upper half of the potential). This difficulty 
can be overcome by loading at higher gradients (\unit[230]{G/cm}) and lowering 
$B_T$ to a final value of \unit[1.7]{G}. A subsequent decrease in the gradient 
over $\sim$ \unit[400]{ms} (limited by speed of power supply) to \unit[80]{G/cm} fully 
splits the cloud (see absorption image of Fig.\ \ref{fig:Loading}). 
Using higher values for $B_{T}$ during this loading procedure decreases 
the barrier height and prevents atoms from staying in the upper well. 
Lower values for $B_{T}$ decrease the coupling at the potential minima and 
result in an insufficient lifetime to observe the separated clouds. The 
chosen values balance these competing effects.

\begin{figure}
\begin{center}
\includegraphics[scale = 1]{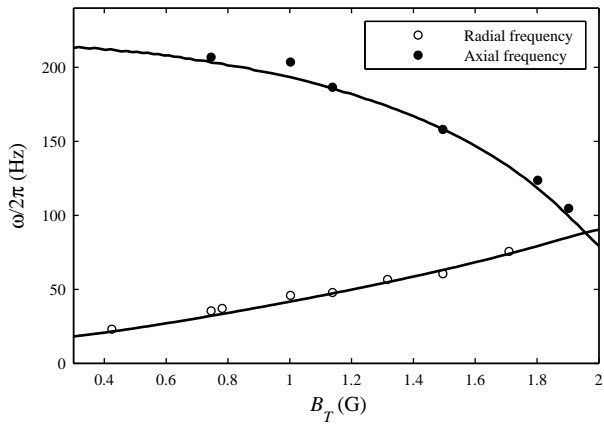}
\end{center}
\begin{flushleft}
\caption[Taap frequency data]{\label{fig:FreqData} Trapping frequencies of the atoms 
in the lower well of the TAAP trap as a function of the magnitude of the rotating 
field $B_{T}$. These frequencies are for a TAAP dressed with circularly polarized RF, 
$B_{RF} =$ \unit[0.5]{G}, and for $B_{q}^{\prime} = $ \unit[84]{G/cm}.}
\end{flushleft}
\end{figure}

In addition the reduced lifetime explains why no data was taken in Fig.\ \ref{fig:PositionData} for 
$B_T <$ \unit[0.4]{G} as the lifetime proved to be insufficient to make reliable measurements of the position. 
This effect is due to Landau-Zener (LZ) transitions to untrapped dressed states. The LZ transition probability 
for transitions between adjacent dressed states in the $F=1$ manifold 
is given by \cite{Vitanov1997}
\begin{equation}
\label{eqn:LZProb}
P_{LZ}({\bf r}) = 1  - \left(1 - \exp\left( -\frac{\pi \hbar \Omega_{R}^2({\bf r})}{2
g_{F}\mu_{B}B_{q}^{\prime}v} \right)\right)^{2},
\end{equation} 
where $v$ is the velocity of the atom through the avoided crossing.
The lifetime $\tau$ of atoms in the TAAP trap varies as 
$\tau \propto 1/P_{LZ}({\bf r}) + \tau_{0}$ where the offset 
$\tau_{0}$ takes into account the finite extent of the 
cloud. Fig.\ \ref{fig:lifetime} shows the variation 
in trap lifetime for a TAAP dressed with linearly polarized 
RF. For this polarization $\Omega_{R}({\bf r})$ is a linear 
function of $B_{T}$ and the lifetime changes by two orders of 
magnitude as $B_{T}$ is ramped down. 
By choosing circularly polarized RF of the appropriate handedness one 
can engineer a situation where the Rabi frequency increases 
as $B_{T}$ is lowered. In this case we apply RF fields in two directions 
each with an amplitude of $B_{RF} =$ \unit[0.5]{G}. The predicted Rabi 
frequency $\Omega_{R} \sim$ \unit[300]{kHz} close to the south pole of the 
ellipsoidal surface agrees well with spectroscopy measurements of the 
trap bottom. These effects combine to give a lifetime of up to \unit[10]{s} in the 
lower TAAP well. In addition, we have measured trapping frequencies as a 
function of $B_{T}$ and observed good agreement with theoretical predictions as 
shown in Fig.\ \ref{fig:FreqData}. In this trap we have successfully 
cooled a thermal cloud to quantum degeneracy. Starting with a sample of 
$1.5\times 10^6$ atoms at $\sim$ \unit[0.7]{$\mu$K} in the TOP trap, we observe an atom loss of 
approximately a third during the TAAP loading process with no substantial heating. The 
loss mechanism is attributed to increased Landau-Zener losses when the avoided crossing 
spirals through the cloud. A subsequent rf-evaporation sweep over \unit[3]{s} with an additional 
weaker field ($\sim$ \unit[0.05]{G}) cools the sample below the critical temperature creating a BEC 
of 5 $\times 10^4$ atoms. For these frequency sweeps we have used transitions 
both within and beyond the rotating-wave approximation (RWA) and observed 
comparable efficiencies \cite{Hofferberth2007}. A BEC can be held in the 
TAAP trap without any evaporative rf for more than \unit[3]{s} without any 
discernible heating.

In conclusion we have presented the first realization of a time-averaged 
adiabatic potential which promises an 
accessible route to a large variety of purely magnetic trapping geometries 
with tunable parameters. The loading scheme presented above 
has acceptable losses and negligible heating. 
Furthermore, we show it is possible to evaporatively cool 
atoms to below quantum degeneracy in such a potential. 
Since completing the work presented above, we have successfully 
employed the TAAP trap as an intermediate stage for loading a radio frequency 
dressed-state potential (lowering $B_T$ to zero) which is 
of interest for studying of low dimensional physics \cite{Zobay2001}. 

This work has been supported by the EPSRC under grant EP/D000440.

\end{document}